# Electronic properties of $A_2Zr_2O_7$ (A= Gd, Nd) ceramic

Algama Masud[1], A. K. Himanshu[2,3,*], Ratnesh Pandey[2], J. Lahiri[2], Nisith Das[2,3], Bijay K Singh[1], Kaustava Bhattacharyya[3,4], Ravi Kumar[3], A. K. Tyagi[3,4]

[1]Department of Physics, TM Bhagalpur University, Bihar-812007, India
[2]Variable Energy Cyclotron Center (VECC), DAE, 1/AF Bidhannagar, Kolkata, 700064, India
[3]Homi Bhabha National Institute, Mumbai, 400094, India
[4]Bhabha Atomic Research Center (BARC), Trombay, Mumbai, 400085, India

**Abstract**

The density functional theory with generalized gradient approximation has been used to investigate the electronic structure of gadolinium pyrochlore $A_2Zr_2O_7$ (A=Gd, Nd) ceramic synthesized in polycrystalline form by solid state reaction. Structural characterization of the compound was done through X-ray diffraction (XRD) followed by Rietveld analysis of the XRD pattern. The Zr-K edge X-ray absorption (XAFS) spectra of $A_2Zr_2O_7$ (A=Gd, Nd) were analysed together with those Zr-foil, which was used as reference compounds. X-ray photoemission spectroscopy (XPS), X-ray absorption near edge structure (XANES) and extended X-ray absorption fine structure (EXAFS) for $A_2Zr_2O_7$ (A=Gd, Nd) has been employed to obtain quantitative structural information on the Zr-local environment. The band gap is estimated using UV-Vis spectroscopy. The crystal structure is face centered cubic, space group being Fd-3m (No. 227). The total energies in this work were calculated using the generalized gradient approximation to DFT plus on-site repulsion (U) method.

## INTRODUCTION

Pyrochlores have a wide range of composition due to their structural flexibility; hence pyrochlores find wide applications such as in nuclear waste management [1, 2]. The pyrochlore structure is a derivative of the ideal fluorite structure. The ordered pyrochlores have the general formula $A_2B_2O_7$, where A is the larger cation and B is the smaller one. The large number of pyrochlore oxides known in the nature are of ($3^+$, $4^+$) type. The pyrochlores belong to Fd-3m space group (No. 227) with $A^{3+}$ cations occupying the 16d, $B^{4+}$ atoms at 16c and oxygen occupying 48f and 8b positions. As all the other atoms occupy special crystallographic sites, the only internal positional parameter that can vary is "$x$" of 48f oxygen atoms. The structure can be viewed as made of eight fold and six fold coordination polyhedral of oxygen atom around A and B cations. For the polyhedral structure $x$ lies in the range 0.3125-0.375. Gadolinium zirconate has a cation size ratio of $r_A/r_B =1.46$, which is on the boundary of the ordered defect-flourite structure, and the structure should be sensitive to the doping of the actinide elements.

In this brief report, we therefore, focus on understanding the electronic properties and local environment of Gd L3-edge along with best-fit theoretical spectrum as revealed through X-ray photoelectron spectroscopy for $A_2Zr_2O_7$ (A= Gd, Nd) .

**Methods**

Sample of $Gd_2Zr_2O_7$ was synthesized using solid-state reaction at 1450 °C for 72 hours. X-ray diffraction was carried out and was analyzed using the Riedvelt analysis software FULLPROF. The structure is found to be cubic, space group Fd-3m (No. 227) with lattice constant a= 10.5249 Å . The positions of Gd, Zr, O1, and O2 atoms are found to be 16d, 16c, 48f and 8b respectively with the *x*-coordinate for the O1 atom in position 48f equal to 0.3456. Calculation of the electronic band structure for this pyrochlore was performed using the Full Potential - Linear Augmented Plane Wave (FP-LAPW) method, in the framework of first principle Density Functional Theory (DFT), as implemented in WIEN2k [3].

Starting with the experimentally obtained structure and atomic positions, volume optimization has been performed to obtain the fully-relaxed structure, obtained from the Birch-Murnaghan Equation of State (BM-EOS) [4]. The value of optimal volume is found to be 2016.9455 a.u.$^3$. We used 72 k-points in the Brillouin zone and the muffin-tin radii for Gd, Zr, O1, and O2 are, respectively, obtained to be 2.35, 2.09, 1.89 and 1.89. The density plane cut-off $R*k_{max}$ is -8.0 Ry, which determines the matrix size (convergence), where $k_{max}$ is the plane-wave cut-off and R is the smallest of all atomic radii. The exchange and correlation effects have been treated within the Generalized Gradient Approximation (GGA). The self-consistency is better than 0.001 e/a.u.$^3$ for charge and spin density, and the stability is better than 0.01 mRy for the total energy per unit cell.

The XAS measurements have been carried out at the Energy-Scanning EXAFS beamline (BL-9) at the Indus-2 Synchrotron Source (2.5 GeV, 100 mA) at Raja Ramanna Centre for Advanced Technology (RRCAT), Indore, India [7, 8]. This beamline operates in the energy range of 4 KeV to 25 KeV. The beamline optics consists of a Rh/Pt coated collimating meridional cylindrical mirror and the collimated beam reflected by the mirror is monochromatized by a Si(111) (2d=6.2709 Å) based double crystal monochromator (DCM). The second crystal of DCM is a sagittal cylinder used for horizontal focusing while a Rh/Pt coated bendable post mirror facing down is used for vertical focusing of the beam at the sample position. Rejection of the higher harmonics content in the X-ray beam is performed by detuning the second crystal of DCM. In the present case, XAS measurements have been performed in both transmission mode and fluorescent mode

### RESULTS AND DISCUSSIONS

In this calculation, we have used the optimum values of the lattice parameters obtained from experimental data. The internal parameters of the atoms have been kept fixed at the experimental values. The optimum values of the lattice parameters are calculated from the ground state energies as a function of unit cell volume. Our calculated results are then fitted using the Birch-Murnaghan Equation of State given by

$$E(V) = E_0 + (9V_0 B_0 /16) \{[(V_0 / V)^{2/3} – 1]^3 B'_0 + [(V_0 / V)^{2/3} – 1]^2 [6 – 4 \{[(V_0 / V)^{2/3}]\} \quad …..(1)$$

The value of the optimal volume is found to be 1046.9829 a.u.$^3$, corresponding to the lattice parameter 10.6116 Å. The DFT calculations were performed with relaxed structure parameters. Fig.1 shows the total energy variation as a function of the unit cell volume per unit formula unit (f.u). The smooth curves are the corresponding fit to the BM-EOS, obtaining the equilibrium structural parameter. The optimal values of the lattice parameters correspond to the atomic positions such that the average force per atoms is less than 1mRy/a.u. The calculated results are then fitted with BM-EOS as Table 1 shows the equilibrium values of the parameters in the BM-EOS.

**Table 1**

|  | EOS | B-M |
|---|---|---|
| Vo | 2015.9465 | 2015.9473 |
| Bulk modulus B (GPa) | 183.8399 | 183.7857 |
| BP | 5.5155 | 5.5163 |
| EO |  | -121148.783645 |

The total density of states of (TDOS) calculated using Generalised Gradient Approximation shows insulating behaviour in the up spin channel and with bandgap of 2.01eV and in the down spin channel with insulating band gap of 2.38eV [5]. In this present study the calculations were performed with U=7eV. The Hubbard U correction was introduced using the method proposed by Duradev et. al [6] in which the parameter U, reflecting the strength of on-site Coulomb interaction, and parameter J, adjusting the strength of exchange interaction are combined into a single parameter $U_{eff} = U-J$. The value of $U_{eff} = 7eV$ (for Gd) has been employed in this calculation in which we get Bohr magnetic moment $\mu_B = 6.96849$. The band structure of the up and down channels is shown in Figs. 1(a) and (b) respectively. It is observed that the direct energy band-gap exists at Γ-point, having insulating nature with a band-gap of approximately 2.19eV (spin up) and 2.26eV (spin down). The valence bands are mainly contributed from the Gd (4f) states. Hybridization of O (2p) with Zr (4d) states is obtained from the Gd (4f) states. The discrepancy between the theoretical and experimental values of band-gap (4.2eV) is due to various approximations involved in the DFT calculations. Other Diffuse Reflectance spectroscopy (DRS), EXAFS, XAS, XANES, XPS detail calculations will be reported elsewhere.

**X-ray photoemission spectroscopy (XPS):**

The XPS spectrum of $Gd_2Zr_2O_7$ & $Nd_2Zr_2O_7$ is shown in Fig. 2a & 2b in the energy window of 0–1100eV. The profiles of the XPS spectra are identified and indexed.

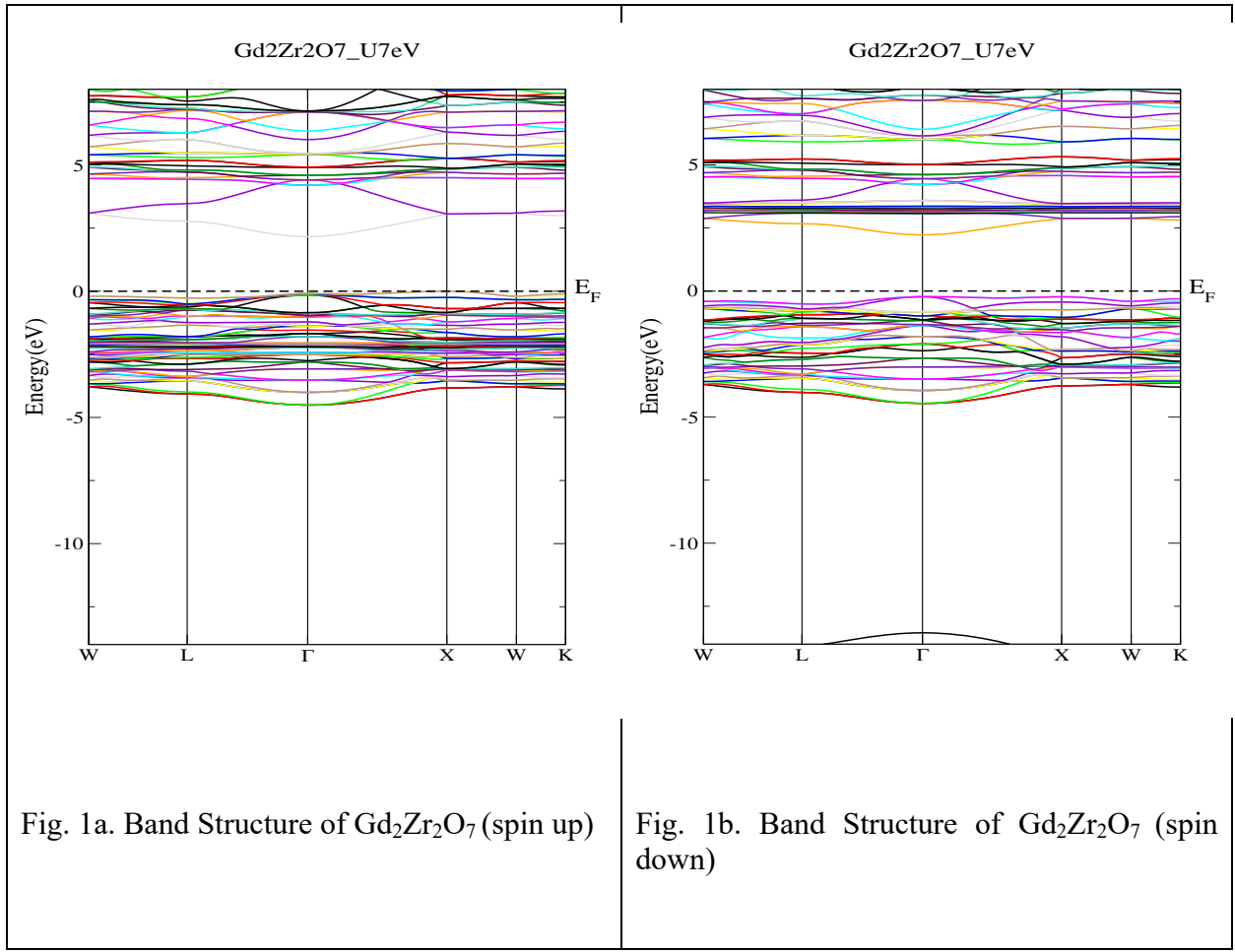

Fig. 1a. Band Structure of $Gd_2Zr_2O_7$ (spin up)

Fig. 1b. Band Structure of $Gd_2Zr_2O_7$ (spin down)

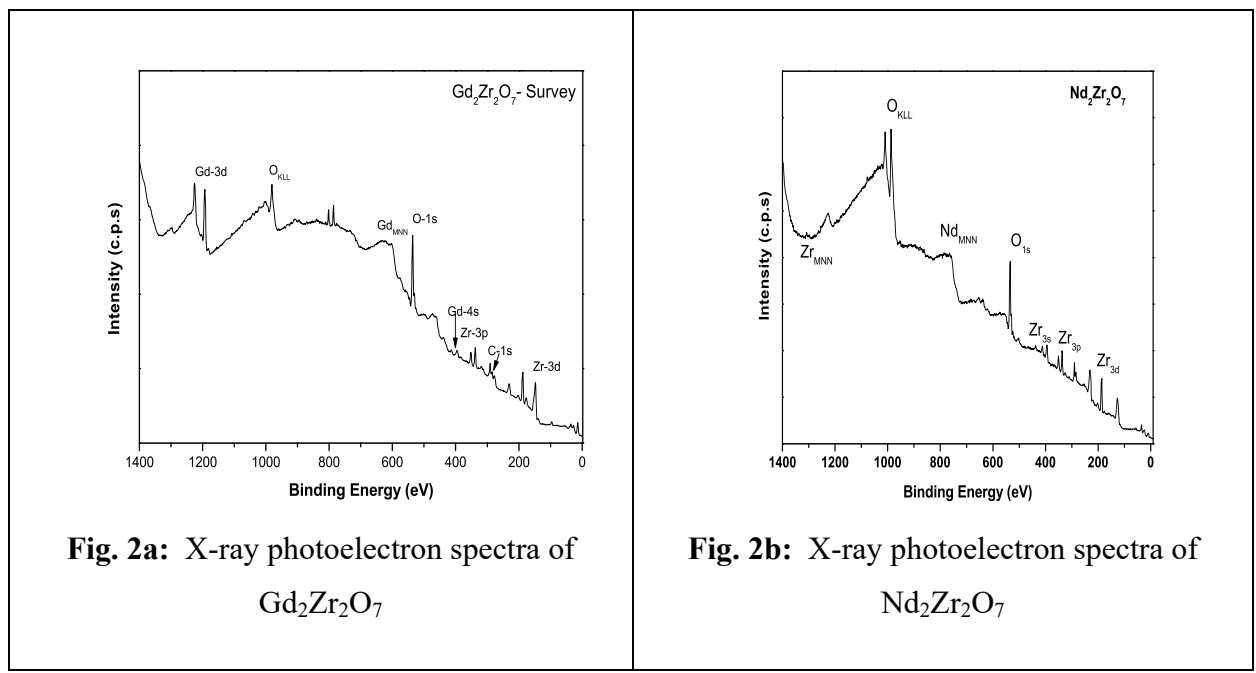

**Fig. 2a:** X-ray photoelectron spectra of Gd$_2$Zr$_2$O$_7$

**Fig. 2b:** X-ray photoelectron spectra of Nd$_2$Zr$_2$O$_7$

The normalized absorption spectrum of the Gd$_2$Zr$_2$O$_7$ sample at Gd L3-edge has been shown in figure 3. The EAXFS data of Gd$_2$Zr$_2$O$_7$ at Gd L3-edge has been fitted by assuming Gd$_2$Zr$_2$O$_7$ crystal structure. The data has been fitted in the range of 1.0-3.5 Å in $R$ space. The peak at 2.0 Å has contributions of Gd-O$_1$ and Gd-O$_2$ coordination shells while the peak at 3.0 Å has contributions of Gd-Zr and Gd-Gd coordination shells. The best fit theoretical plot has been shown in fig. 4 alongwith the experimentally derived $\chi(r)$ vs $r$ .data and the best fit parameters have been shown in table 2.

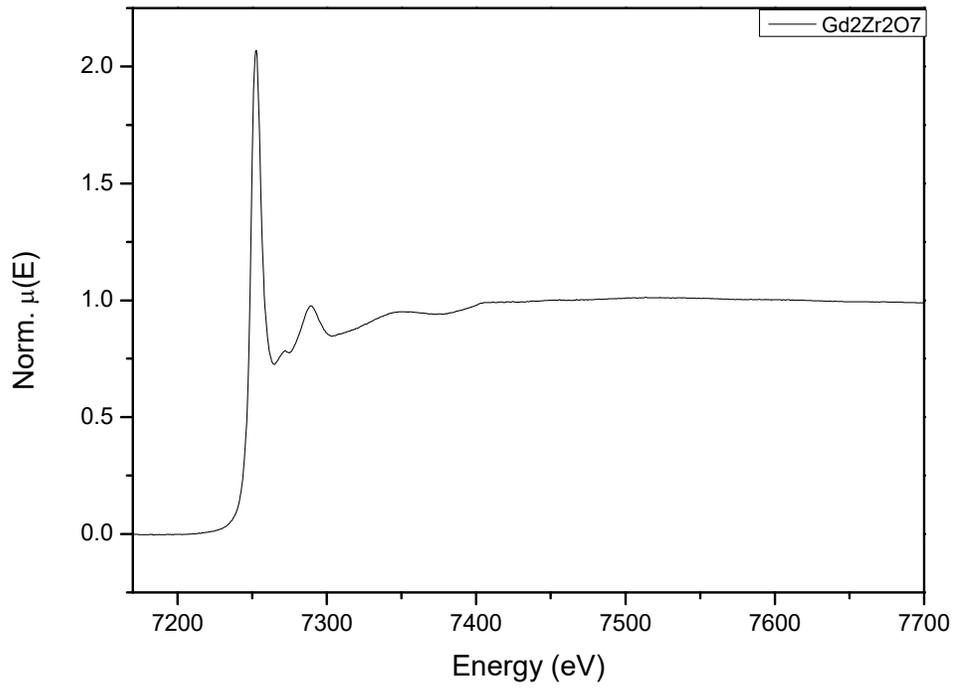

**Figure 3** Normalized absorption spectrum of $Gd_2Zr_2O_7$ sample measured at Gd L3-edge.

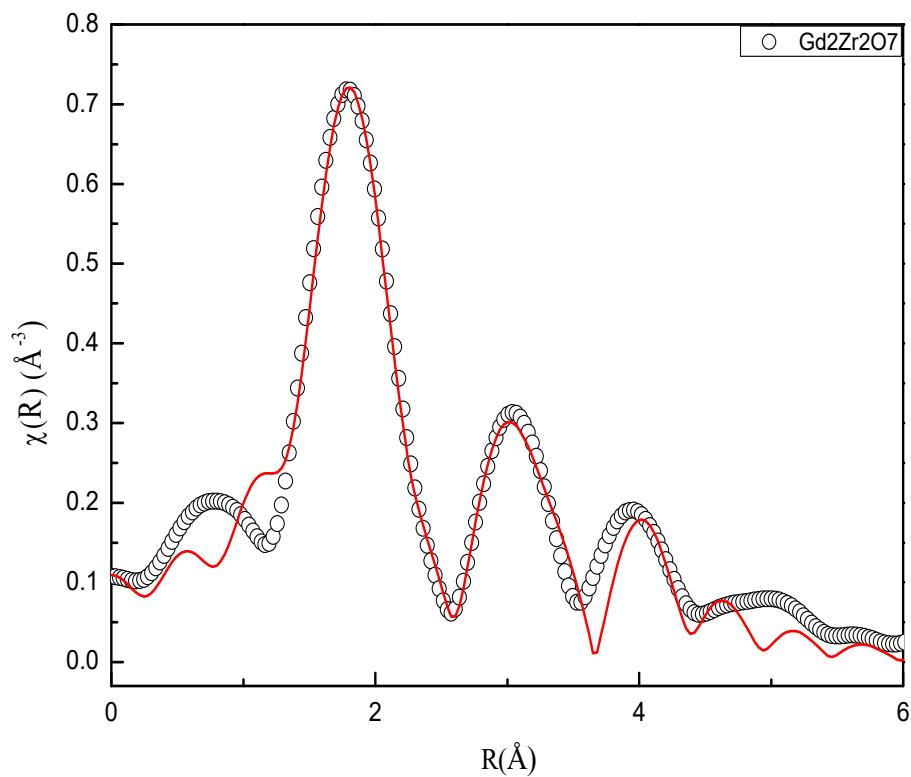

**Fig. 4** Fourier transformed EXAFS spectrum of $Gd_2Zr_2O_7$ sample measured at Gd L3-edge alongwith best-fit theoretical spectrum. The experimental spectrum is represented by scatter points and theoretical fit is represented by solid line.

**Table 2** Bond length, coordination number and disorder factors obtained by EXAFS fitting for $Gd_2Zr_2O_7$ sample measured at Gd L3-edge.

| Path | Parameter | $Gd_2Zr_2O_7$ |
|---|---|---|
| | Gd edge | |
| Gd-$O_1$ | R (Å) (2.28) | 2.31±0.005 |
| | N(2) | 2±0.06 |
| | $\sigma^2$ | 0.0031±0.0008 |
| Gd-$O_2$ | R (Å) (2.48) | 2.46±0.007 |
| | N(6) | 6±0.18 |
| | $\sigma^2$ | 0.03±0.0017 |
| Gd-Zr | R (Å) (3.73) | 3.64±0.008 |
| | N(6) | 6±0.18 |
| | $\sigma^2$ | 0.0116±0.0011 |
| Gd-Gd | R (Å) (3.73) | 3.70±0.007 |
| | N(6) | 6±0.18 |
| | $\sigma^2$ | 0.0065±0.001 |
| $R_{factor}$ | | 0.01 |